\documentclass[conference]{IEEEtran}

\usepackage{graphicx}
\usepackage{subfigure}
\usepackage{cite}
\usepackage{bm,bbm}
\usepackage{algorithm,algpseudocode}

\usepackage{booktabs}

\graphicspath{{./figs/}}
\usepackage{url}

\usepackage{amsmath,amssymb,amsfonts,mathtools,amsthm}

\usepackage{comment}
\usepackage{subfiles}

\theoremstyle{definition}
\theoremstyle{definition}

\ifCLASSOPTIONcompsoc
	\usepackage[font=normalsize,labelfont=sf,textfont=sf]{subfig}
\else
	\usepackage[font=footnotesize]{subfig}
\fi

\begin{document}
\title{Adversarial Reinforcement Learning-based \\Robust Access Point Coordination \\Against Uncoordinated Interference}
\author{
\IEEEauthorblockN{
\normalsize Yuto Kihira\IEEEauthorrefmark{1},
\normalsize Yusuke Koda\IEEEauthorrefmark{1}\IEEEauthorrefmark{2},
\normalsize Koji Yamamoto\IEEEauthorrefmark{1}\IEEEauthorrefmark{3},
\normalsize Takayuki Nishio\IEEEauthorrefmark{1}, and
\normalsize Masahiro Morikura\IEEEauthorrefmark{1}
}
\IEEEauthorblockA{
\IEEEauthorrefmark{1}\small Graduate School of Informatics, Kyoto University,
Yoshida-honmachi, Sakyo-ku, Kyoto 606-8501, Japan\\
\IEEEauthorrefmark{2}\small koda@imc.cce.i.kyoto-u.ac.jp, 
\IEEEauthorrefmark{3}\small kyamamot@i.kyoto-u.ac.jp 
}
}

\maketitle
\begin{abstract}
	This paper proposes a robust adversarial reinforcement learning (RARL)-based multi-access point (AP) coordination method
	that is robust even against unexpected decentralized operations of uncoordinated APs.
	Multi-AP coordination is a promising technique towards IEEE 802.11be,
	and there are studies that use RL for multi-AP coordination.
	Indeed, a simple RL-based multi-AP coordination method diminishes the collision probability among the APs;
	therefore, the method is a promising approach to improve time-resource efficiency.
	However, this method is vulnerable to frame transmissions of uncoordinated APs
	that are less aware of frame transmissions of other coordinated APs.
	To help the central agent experience even such unexpected frame transmissions,
	in addition to the central agent,
	the proposed method also competitively trains an adversarial AP that disturbs coordinated APs by causing frame collisions intensively.
	Besides, we propose to exploit a history of frame losses of a coordinated AP
	to promote reasonable competition between the central agent and adversarial AP.
	The simulation results indicate that the proposed method can avoid uncoordinated interference
	and thereby improve the minimum sum of the throughputs in the system compared to not considering the uncoordinated AP.
\end{abstract}
\IEEEpeerreviewmaketitle

\section{Intoroduction}
\label{sec:Introduction}
The popularity of 802.11 wireless LANs (WLANs) is growing rapidly
owing to their ease of deployment, convenience, and cost efficiency.
The rapid spread of WLANs has created dense deployments of WLAN devices,
which must compete for the limited transmission opportunities.
In these environments, carrier sense multiple access with collision avoidance (CSMA/CA) mechanism,
which is a distributed control mechanism, causes frequent frame losses and results in considerable throughput degradation
because each WLAN access point (AP) can experience severe inter-cell interference from neighboring WLAN APs.

To improve the spectrum efficiency in high density WLANs, a centralized coordination of APs has attracted increased attention.
For example, in the industry, the application of coordinated orthogonal frequency division multiple access (OFDMA)
is under discussion for IEEE 802.11be standard\cite{lopez2019ieee}, which is being standardized for the next-generation extremely high throughput WLANs, to eliminate frame collisions completely.
In academics, by exploiting a connection among APs and a central component through a wired backhaul\cite{zhu2008multi},
a scheme determining the clear channel assessment threshold\cite{nakahira2014centralized}
and a scheme allocating channels to the APs\cite{lim2016centralized} have both been proposed.
The trend in multi-AP coordination includes reinforcement learning (RL)-based approaches\cite{nakashima2019deep},
which improves throughput by allocating different channels to APs within the carrier sensing range of each AP.

However, not all APs are necessarily under the coordination of a central agent
when an AP operated in a distributed manner is placed in the proximity to the coordinated APs.
If such an uncoordinated AP is operated around the APs under the coordination based on a learned control policy,
a coordinator may not be able to make a reasonable decision
because most RL-based methods exhibit a poor performance
when an agent is affected by a disturbance that is not experienced previously.

Focusing on an AP coordination scenarios as an example,
we propose incorporating a recent robust reinforcement learning framework called robust adversarial reinforcement learning (RARL)\cite{RARL}.
In the proposed RARL-based AP coordination, a coordinator learns a transmission policy
whereas an uncoordinated AP with an intelligence termed as an adversarial AP trains to disturb a coordinated AP by transmitting a frame
and thereby causing a frame collision.
This competitive training enables the coordinator to experience various frame losses incurred by the uncoordinated AP,
and through such experiences, the coordinator assigns time slots so that the frame losses can be avoided.
Due to such various experiences, this learned policy exhibits a desired performance without relearning
even if environment changes (e.g., an uncoordinated AP is suddenly placed or the transmission probability of the uncoordinated AP changes).
To the best of our knowledge,
no prior studies have considered the use of RARL in the context of enhancing the robustness about WLANs.
\begin{figure*}[!tb]
    \centering
    \subfigure[Test scenario.]{%
        \includegraphics[width=1.0\columnwidth]{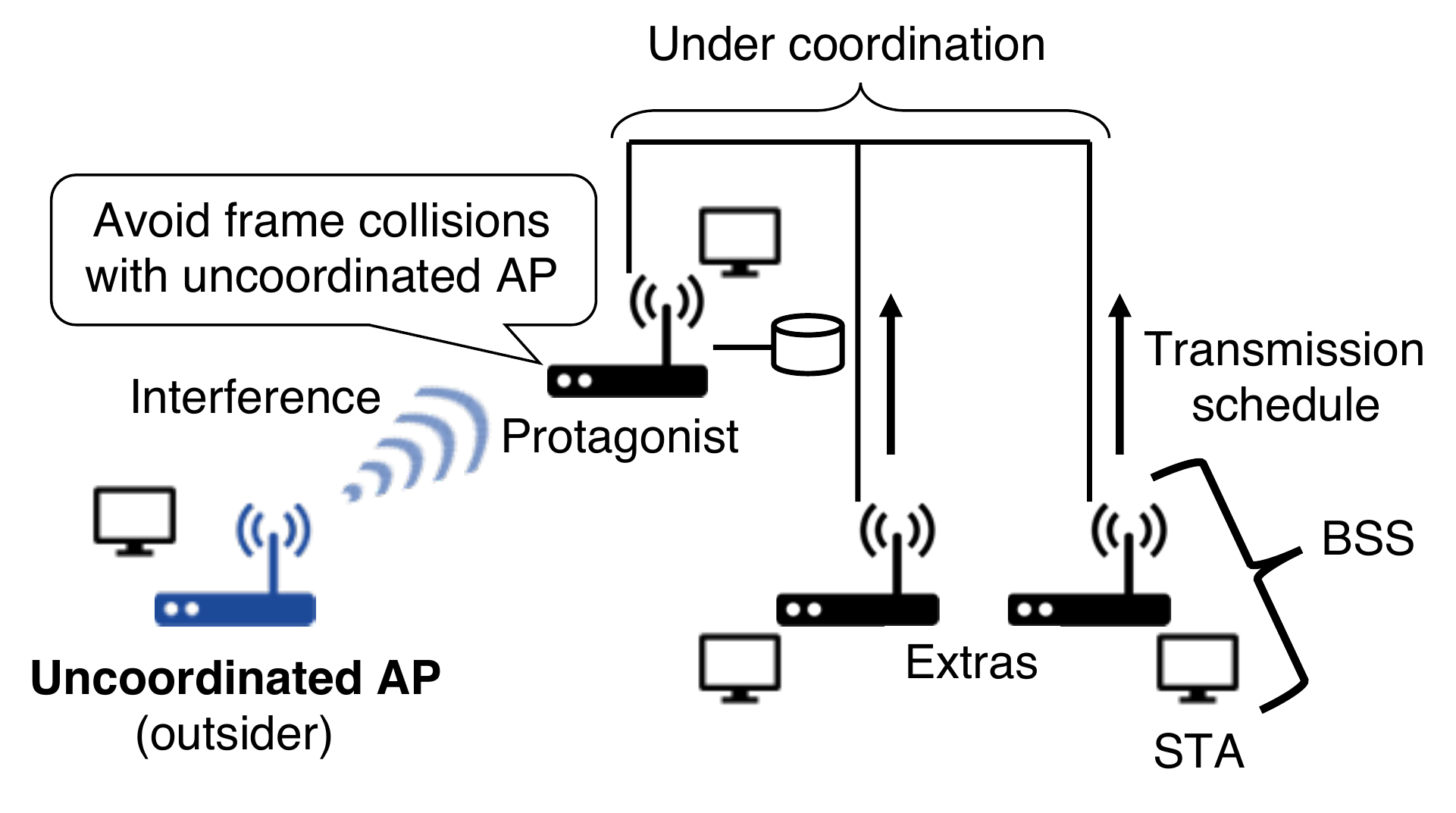}}
    \subfigure[Training scenario.]{
        \includegraphics[width=1.0\columnwidth]{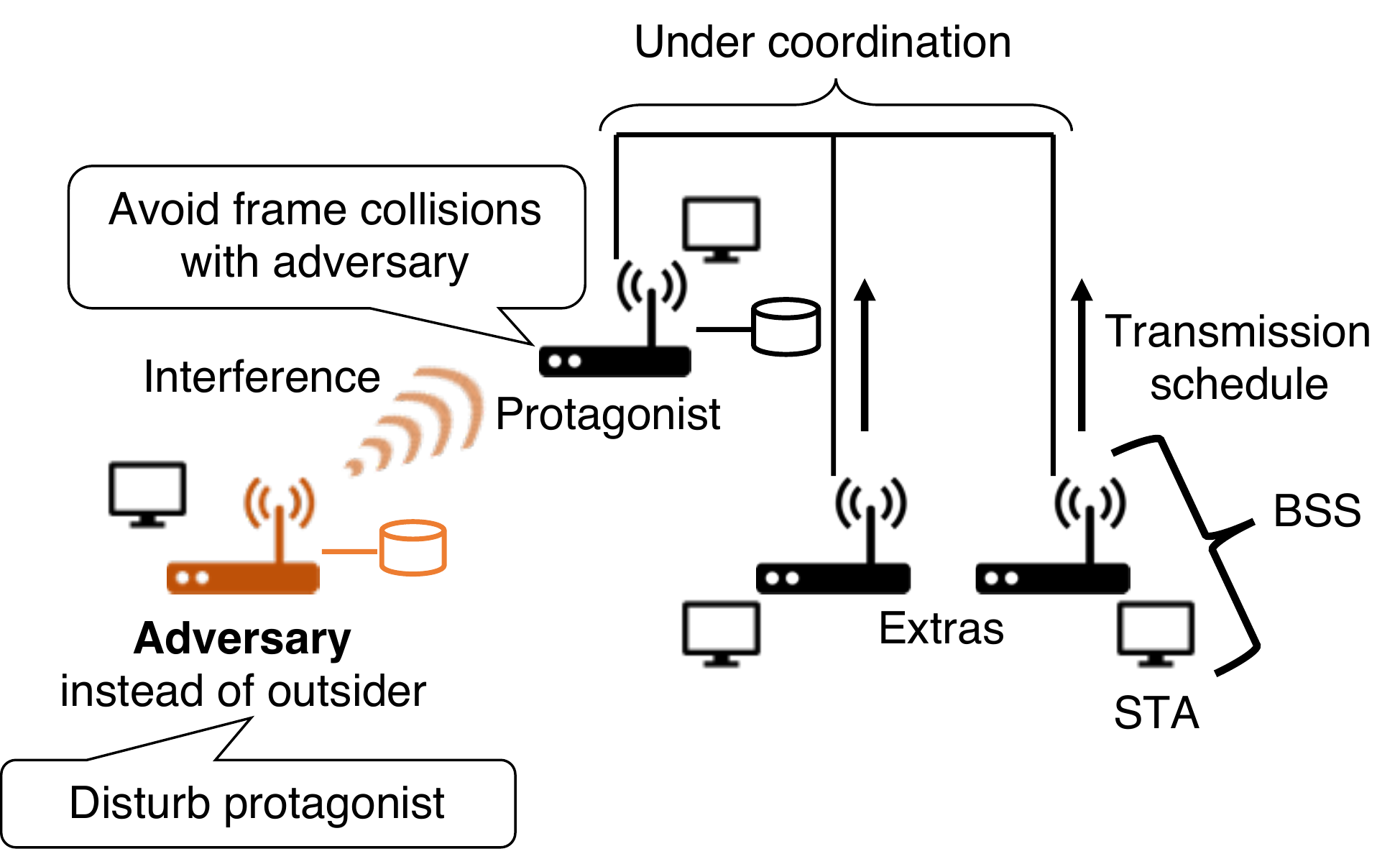}}
    \caption{Overview of the proposed scheme.}
    \label{fig:overview}
\end{figure*}

The main contribution of this paper is as follows:
We propose an RL-based multi-AP coordination method that is robust against even unexpected operations of uncoordinated APs,
which is based on the following two key ideas.
First, we apply RARL, where the central agent competitively learns a frame transmission policy with an adversarial AP
that learns a policy to disturb a coordinated AP by causing frame collisions.
By experiencing such disturbances with the help of the adversarial AP, the central agent can become robust to frame transmissions of uncoordinated APs.
However, only training the original agent with the adversarial AP (i.e. straightforward application of RARL) is not sufficient to encourage the original agent to learn a robust policy.
This is because the original agent needs to predict and oppose the actions of the adversarial agent
although the original agent does not obtain sufficient information about the tendency of the actions.
Hence, to inform the original agent of the action preference of the adversary and thereby to encourage the robustness of the policy,
the second idea is to utilize a history of frame losses of the typical AP
and to design a reward that depends on a set of actions of both agents.

The rest of this paper is constructed as follows.
Section II describes our system model.
Section III describes both the proposed and the competing algorithms.
Section IV presents the simulation evaluations of the proposed scheme in comparison to the other schemes.
Section VI provides some concluding remarks regarding this paper.

\section{System Model}
\label{sec:SystemModel}
\begin{figure}[tb]
    \centering
    \includegraphics[width=1\columnwidth]{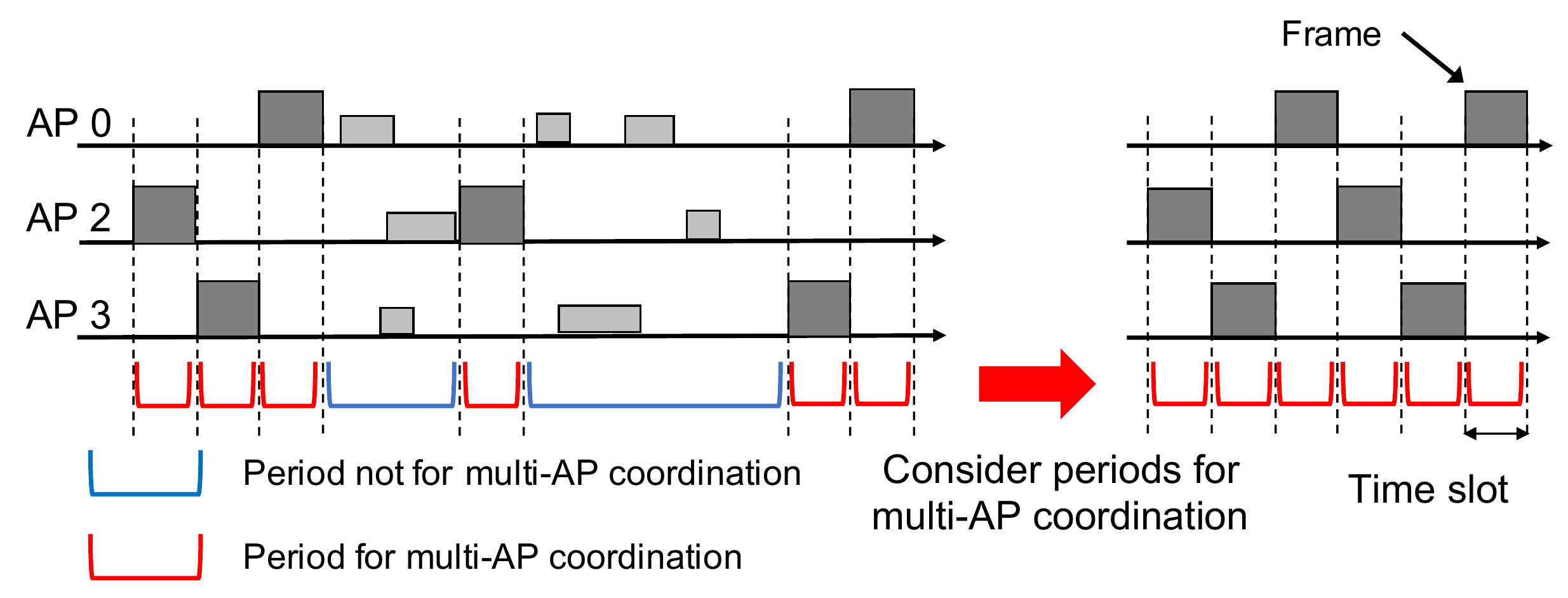}
    \caption{Considered coordinated periods.}
    \label{fig:extraction}
\end{figure}

The system model is shown in Fig.~\ref{fig:overview}(a).
Assume that there are coordinated $N$ basic service sets (BSSs),
and for the ease of explanation, each of them consists of an AP and a station (STA) with only downlink traffic.
Note that this assumption can be easily extended to multiple STA scenario.
These BSSs perform coordinated time-division resource assignment in the same frequency band in periods for multi-AP coordination,
which is under discussion for 802.11be standard\cite{lopez2019ieee}.
In this paper, the unit of resource assignment is called as a slot as shown in  Fig.\ref{fig:extraction}.
AP~0 (called the protagonist) is a typical and intelligent agent.
The other APs (called extras) share information on their transmission schedules with the protagonist.
AP~1 (called the outsider) is an uncoordinated AP,
i.e., it does not share information on its transmission schedule with the protagonist
and thus can prevent successful frame transmissions of the protagonist owing to the frame collision.
Note that we only consider slots for multi-AP coordination.
In other words, we already delete periods during which contention-based procedures are needed as shown in Fig.~\ref{fig:extraction}.

The protagonist determines whether to transmit a frame, as well as the transmission data rate,
(i.e., a modulation and coding scheme (MCS) index)
according to the information on the transmission schedules of the extras.
It is worth noting that, because as all APs transmit under the same power,
the signal-to-interference-plus-noise ratio (SINR) at STA~0 decreases
under the condition in which some extras or an outsider are transmitting a frame.
Thus, the transmission of the protagonist is not necessarily successful under such conditions.
After the protagonist finishes transmitting a frame,
STA 0 immediately sends an acknowledgment frame upon the successful repetition of the transmitted frame\cite{handbook},
and thus, the protagonist knows whether the frame transmission is successful.

\section{Adversarial Reinforcement Learning for Learning Robust Policy to Uncoordinated AP}
\label{sec:LeaningProcess}
The objective of this study is to learn a robust policy against APs that do not share the information on their transmission schedules.
To this end, we propose an RARL-based scheme.
The overview of this scheme is shown in Fig.~\ref{fig:overview}.
In the proposed scheme, the protagonist learns the transmission policy along with a hypothetical adversarial AP (called the adversary) regarding the training process.
The adversarial AP learns a transmission policy that effectively disturbs the frame transmission of the protagonist.
The protagonist learns a countermeasure policy and becomes robust to the disturbance from the adversarial AP.
Thus, the protagonist is expected to exhibit robustness to uncoordinated APs under the test scenario
because the protagonist has experienced a more intelligent disturbance from the adversarial AP under the training scenario.


\subsection{Markov Game for Jointly Learning Transmission Policy of Protagonist and Adversarial AP}
Under the training scenario, we consider an adversarial setting in which the protagonist and adversary compete,
which is expressed as a two-player $\gamma$ discounted zero-sum Markov game\cite{littman1994markov,scherrer2015approximate}.
This Markov game can be expressed as the tuple
$(\mathcal{S}, \mathcal{A}_0, \mathcal{A}_1, \mathcal{T}, r, \gamma, P)$,
where $\mathcal{S}$ denotes the state space,
$\mathcal{A}_0$ denotes the set of possible actions of the protagonist,
$\mathcal{A}_1$ denotes the set of possible actions of the adversary,
$\mathcal{T} : \mathcal{S} \times \mathcal{A}_0 \times \mathcal{A}_1 \times \mathcal{S} \rightarrow \mathbb{R}$ denotes the transition probability to the next state,
$r : \mathcal{S} \times \mathcal{A}_0 \times \mathcal{A}_1 \rightarrow \mathbb{R}$ is the reward of both players,
$\gamma$ is the discount factor, and $P$ is the initial state distribution.


The protagonist and adversary observe the same state at each learning step.
The state comprises the following two parts:
(1) the frame transmission states of the extras                                     
and (2) the last $X$ step histories of the transmissions of the protagonist and the adversary.
Thus, the state space is given as follows:
\begin{align}
    \mathcal{S} \coloneqq \mathcal{S}_2 \times \mathcal{S}_3 \times\dots\times
    \mathcal{S}_N \times \mathcal{S}_\mathrm{history}.
\end{align}
Here, $\mathcal{S}_i\,(i=2,3,\cdots N-1)$ is given as follows:
\begin{align}
    \mathcal{S}_i \coloneqq \{-1,1\},
\end{align}
where $-1$ and 1 donote AP~$i$ that do not and do transmit, respectivery.           
In addition, $\mathcal{S}_{\mathrm{history}}$ is given as follows:
\begin{multline}
    \mathcal{S}_{\mathrm{history}} \coloneqq \{\,(x_1, x_2, x_3)  \mid x_1, x_2, x_3 \in \mathbb{N}_0 \\
    {} \land x_1+ x_2 + x_3 = X \,\} ,
\end{multline}
where $\mathbb{N}_0$ is non negative integer, and $x_1$, $x_2$, and $x_3$ are the numbers of steps at which the agents take the corresponding actions during the last $X$ steps.
In particular, $x_1$ is the number of steps during which both agents transmit,
$x_2$ is the number of steps during which the protagonist transmits and the adversary does not,
and $x_3$ is the number of steps at which the protagonist does not transmit regardless of the adversary's action.

At each step, the protagonist decides whether to transmit and if so, it selects an MCS index.
Thus, the action space of the protagonist is given as follows:
\begin{align}
    \mathcal{A}_0 \coloneqq \{-1\} \cap \mathcal{M},
\end{align}
where $\mathcal{M}$ denotes a set of MCS indexes and $-1$ denotes an action in which no frames are transmitted.
A set of MCS indexes is given as $\mathcal{M} = \{0,1,\dots,M\}$,
and the respective number denotes which MCS index the protagonist selects when transmitting a frame.
The adversary decides whether to transmit.
The action space of the adversary is given as follows:
\begin{align}
    \mathcal{A}_1 \coloneqq \{-1,1\},
\end{align}
where $-1$ and 1 denote the action of not transmitting any frames and that of transmitting a frame, respectively.
The protagonist's reward $r$ consists of the following two components and is given as follows:
\begin{align}
    r \coloneqq r_1 + r_2,
\end{align}
where $r_1$ is the amount of data of the successful transmission of the protagonist,
and $r_2$ is the additional reward given
if the protagonist can expect the frame transmission of the adversarial AP and is determined as follows:
\begin{align}
    r_2 \coloneqq 
    \begin{cases}
        c, & a^0 = -1 \land a^1 = -1; \\
        -c, & a^0 = -1 \land a^1 = 1; \\
        0, & a^0 \neq -1,
    \end{cases}
    \label{eq:reward}
\end{align}
where $a^0 \in \mathcal{A}_0$ denotes an action of the protagonist,
$a^1 \in \mathcal{A}_1$ denotes an action of the adversary,
and $c$ is a positive constant number.
This additional reward $r_2$ encourages the adversary to take various actions.
Because $r$ depends on the action of an uncoordinated AP,
the protagonist is expected to learn a policy
that also avoids a collision with an uncoordinated AP during the test scenario.
In contrast, the adversary's reward $r'$ is a sign inversion of $r$,
where the adversary learns the transmission policy that minimizes the reward of the protagonist.



\subsection{Learning Algorithm}
The protagonist and adversarial AP learn policies alternately during the training scenario.
First, the protagonist updates its policy, whereas the adversary fixes its policy.
Next, the protagonist's policy is held constant, and the adversary updates its policy.
This process is iterated until the $N_{\mathrm{iter}}$th step is completed.

Algorithm~\ref{alg:RARL} shows the learning procedure.
The part of the initial state that depends on the last $X$ step histories of the transmissions of the protagonist and the adversary
$s_{0,\mathrm{history}} \in \mathcal{S}_{\mathrm{history}}$ is sampled from a uniformly random distribution.
During each iteration, we carry out a two-step (alternating) optimization procedure
and optimize the action value function of the protagonist $Q^{\pi}_0(s,a)$.
The action value function $Q(s,a)$ is defined as the value of taking action $a$ under state $s$ based on policy $\pi$ as follows:
\begin{align}
    Q^{\pi}(s,a) = \mathbb{E}^{\pi}\!\left[\sum^{\infty}_{t=0} \gamma^t r_{t+1}\,\middle|\,s_0 = s, a_0 = a\right].
\end{align}
First, for $N_0$ iterations, the action value function of the adversary $Q^{\pi}_1(s,a)$ is held constant,
whereas the action value function of the protagonist $Q^{\pi}_0(s,a)$ is optimized to maximize the following:
\begin{align}
    R = \mathbb{E} \left[ \sum_{k=0}^{\infty} \gamma^k r_k \right],
\end{align}
where $r_k$ denotes a reward in which the protagonist observes $k$ steps later.
At each step, the protagonist updates its action value function based on the Q-learning method\cite{RL},
\begin{align}
    Q^{\pi}_0(s,a) \leftarrow Q^{\pi}_0(s,a) + \alpha [r + \gamma \max_{a \in \mathcal{A}_0} Q^{\pi}_0(s',a) - Q^{\pi}_0(s,a)].
\end{align}

For the second step, the action value function of the protagonist is held constant for the next $N_1$ iterations.
To minimize $R$, the adversary updates its action value function as follows:
\begin{align}
    Q^{\pi}_1(s,a) \leftarrow Q^{\pi}_1(s,a) + \alpha [- r + \gamma \max_{a \in \mathcal{A}_1} Q^{\pi}_1(s',a) - Q^{\pi}_1(s,a)].
\end{align}
This alternating procedure is repeated for $N_{\mathrm{iter}}$ iterations.


\begin{figure}[!t]
    \vspace{-3mm}
    \makebox[\linewidth]{
        \begin{minipage}{\linewidth}
            \begin{algorithm}[H]
                \floatname{algorithm}{Algorithm}
                \caption {Adversarial reinforcement learning for learning policy robust to uncoordinated AP}
                \label{alg:RARL}
                \begin{algorithmic}[1]
                    \Require Environment $\mathcal{E}$
                    \State \textbf{Initialize:} $Q^{\pi}_0$, $Q^{\pi}_1$ and the initial state $s_0$
                    \For {$i=1,2,\dots, N_\mathrm{iter}$}
                        \For {$j=1,2,\dots, N_0$}
                            \State Choose transmission rate $a^0$ based on $Q^{\pi}_0(s,a)$
                            \State Choose to transmit or not $a^1$ based on $Q^{\pi}_1(s,a)$
                            \State Observe transmission success and reward $r$
                            \State Store a set of actions as history
                            \State Observe next state $s'$
                            \State Update $Q^{\pi}_0(s,a)$ and $s \leftarrow s'$
                        \EndFor
                        \For {$j=1,2,\dots, N_1$}
                            \State Choose transmission rate $a^0$ based on $Q^{\pi}_0(s,a)$
                            \State Choose to transmit or not $a^1$ based on $Q^{\pi}_1(s,a)$
                            \State Observe transmission success and reward $r$
                            \State Store a set of the actions as history
                            \State Observe next state $s'$
                            \State Update $Q^{\pi}_1(s,a)$ and $s \leftarrow s'$
                        \EndFor
                    \EndFor
                \end{algorithmic}
            \end{algorithm}
        \end{minipage}
    }
\end{figure}

The protagonist takes stochastic actions according to \textit{softmax} action selection rules\cite{RL}
when the agents choose actions from the action value functions.
According to the \textit{softmax} action selection, the agents choose actions with the following probability:
\begin{align}
    \frac{\mathrm{e}^{Q(b)/\tau}}{\sum_{b-1}^n \mathrm{e}^{Q(b)/\tau}},
\end{align}
where $\tau$ is a positive parameter called the temperature,                        
and $Q(a)$ is an action value function.

\section{Simulation Evaluation}
\label{sec:SimulationEvaluation}

\subsection{Simulation Settings}
In this section, we confirm the effect of the proposed scheme
on the sum of the throughputs of the outsider and the protagonist
and on the transmission probability of the protagonist.
To make it easier to understand this effect, we compare our proposed scheme with the following two baselines:
\begin{itemize}
    \item Oracle scheme\par
    The protagonist can obtain information on the transmission schedules of all APs including the outsider.
    The throughput in this baseline is the upper limit.
    \item Without an adversarial RL scheme\par
    The protagonist learns a policy without an adversary
    and transmits regardless of collisions against the transmission of the outsider.
\end{itemize}

The simulation parameters are shown in Table~\ref{tbl:SimulationParameters}.
A set of MCS indexes from which the protagonist can select is given as follows:
\begin{align}
    \mathcal{M} = \{1,2,3\},
\end{align}
where 1 denotes a transmission at 1\,Mbit/slot,
2 denotes a transmission at 2\,Mbit/slot,
and 3 denotes a transmission at 3\,Mbit/slot.
To improve readability, we omit index 0.
The transmission rate of the outsider is held 3\,Mbit/slot.
We assume that the positions of all APs are fixed
so that the levels at which the transmissions of each AP interfere with the transmissions of the protagonist are fixed, respectively.
In the training process, by selecting the suitable transmission rate according to the interference levels from extras, 
the protagonist learns a policy that avoids frame collisions with the extras.
In the proposed scheme, the protagonist and adversary learn alternately every 100 time slots under the training scenario.
\begin{table}[tb]
    \centering
    \caption{Simulation parameters.}
    \begin{tabular}{ll}
        \toprule
        Parameters & Values\\
        \midrule
         Number of the extras $N$ & 2\\
         Number of the training iteration $N_\mathrm{iter}$ & $100,000$\\ 
         Slot number of the protagonist's training at a iteration $N_0$ & 100\\
         Slot number of the adversary's training at a iteration $N_1$ & 100\\
         Number of stored slots $X$ & 20\\
         Learning rate $\alpha$ & 0.1\\
         Discount factor $\gamma$ & 0.9\\
         Temperature in training $\tau_1$ & 5\\
         Temperature in test $\tau_2$ & 0.1\\
        \bottomrule       
    \end{tabular}
    \label{tbl:SimulationParameters}
\end{table}

\subsection{Results and Discussions}
\begin{figure}[tb]
    \centering
    \includegraphics[width=1.0\columnwidth]{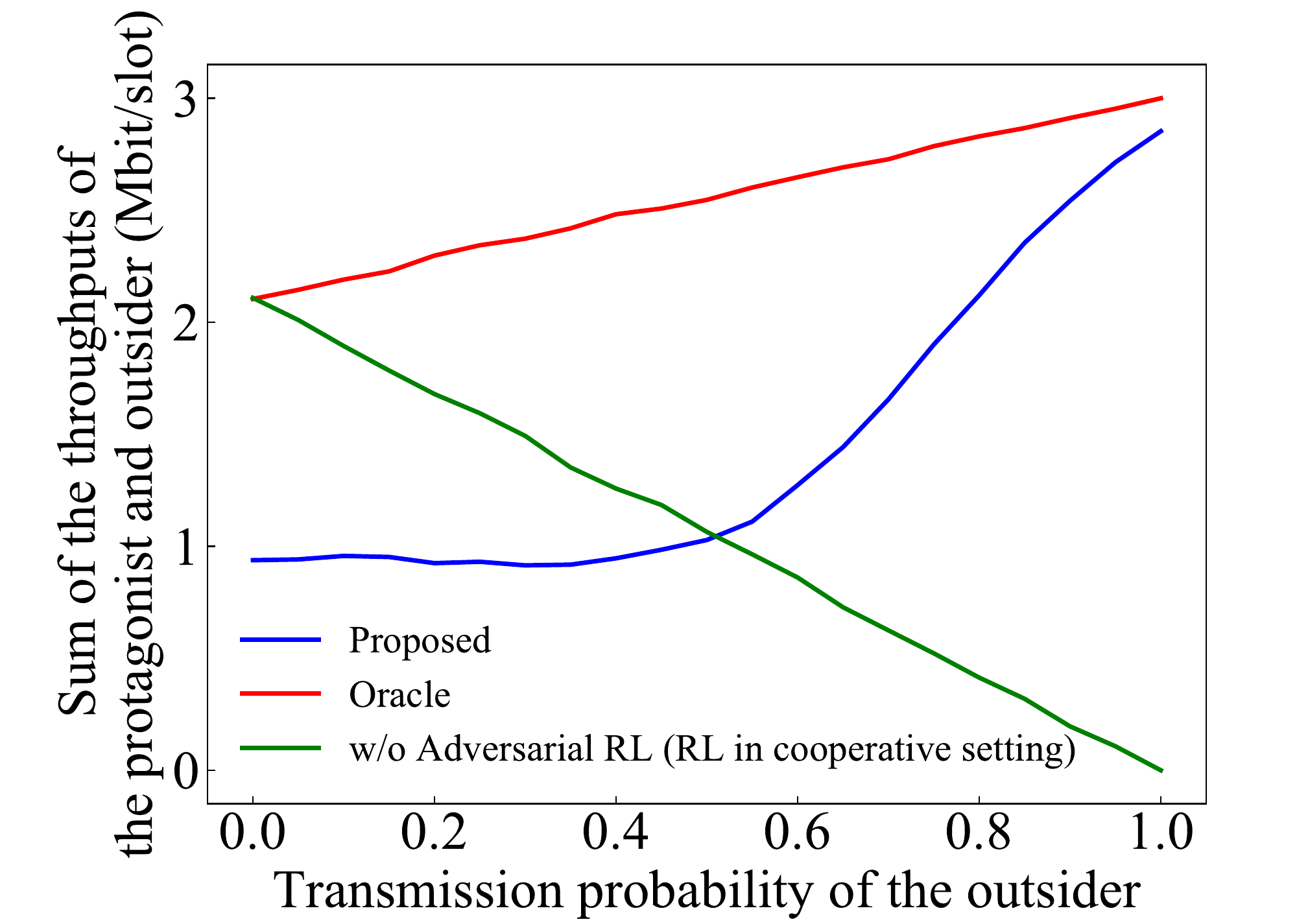}
    \caption{Sum of throughputs of the protagonist and adversary.}
    \label{fig:throughput}
\end{figure}

The relation between the transmission probability of the outsider and the sum of the throughputs
of the protagonist and outsider is shown in Fig.~\ref{fig:throughput}.
As shown in the figure, when the transmission probability of the outsider is high,
in terms of the sum of the throughputs,
the protagonist learning with the adversary under the training scenario is superior to
the one that learns a policy only in a completely cooperative setting.
However, when the transmission probability of the outsider is low, the proposed scheme is inferior.

\begin{figure}[tb]
    \centering
    \includegraphics[width=1.0\columnwidth]{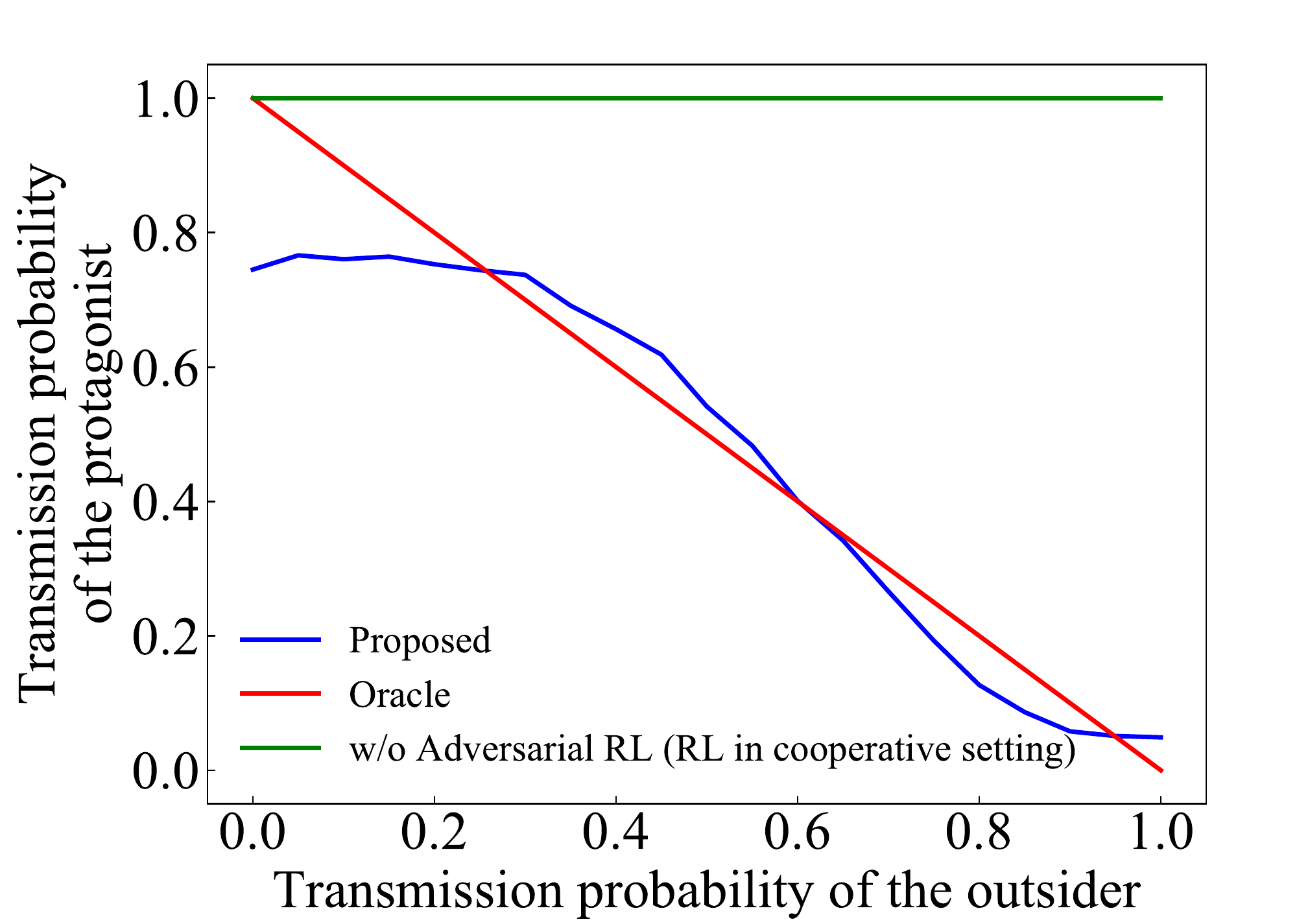}
    \caption{Transmission probability.}
    \label{fig:transmisiion_probability}
\end{figure}

According to Fig.~\ref{fig:throughput}, by using the proposed scheme,
the protagonist maintains the performance with arbitrary transmission probabilities of the outsider.
This is because the protagonist changes its transmission probability according to the transmission probability of the outsider,
which the protagonist cannot observe, as shown in Fig.~\ref{fig:transmisiion_probability}. 
In this sense, the protagonist learns a policy that is robust to changes in the outsider's transmission probability.
By contrast, if the protagonist learns a policy without competing with its adversaries,
the throughput is decreased significantly when the outsider frequently transmits.
This means that the protagonist that does not consider uncoordinated APs
is vulnerable to a change in the transmission probability of the outsider.
 
However, the protagonist using the proposed scheme does not improve the sum of the throughputs compared to without an adversarial RL scheme
when the outsider rarely transmits.
This is because the adversary prefers severe situations for the protagonist under the training scenario,
the adversary tends to transmit a frame to disturb the transmission of the protagonist.
The protagonist has not experienced a condition in which the channel is not crowded with an unexpected transmission,
and finishes the training before learning a sufficient amount under such a condition.
Therefore, even if the protagonist does not lose frames during recent slots and predicts that the transmission probability of the outsider is low,
the protagonist does not increase the transmission probability.

By contrast, the protagonist that determines the actions in accordance with the policy learned without an adversarial RL scheme
is unafraid to lose its frames by a collision with the outsiders' frames.
As a result, the transmission probability is almost 100\% regardless of the outsider, as shown in Fig.~\ref{fig:transmisiion_probability}.
When the outsider seldom transmits, the sum of the throughput is higher than that of the proposed scheme.

As shown in Fig.~\ref{fig:transmisiion_probability}, these probabilities have a negative correlation when the proposed scheme or an oracle scheme is adopted,
whereas the probability of the protagonist takes a constant value when a scheme without an adversarial RL is adopted.
Besides,
the protagonist using the proposed scheme may try to change its transmission probability to follow the line-showing oracle scheme.
However, when the probability of the outsider is low, the line of the proposed scheme deviates,
the reason for which is expected to be the same as described above.
The protagonist has not experienced an uncrowded channel owing to the preference of the adversary.

\section{Conclusion}
\label{sec:Conclusion}
In this study, we presented an RARL-based multi-AP coordination
that is robust against even unexpected decentralized operations of uncoordinated APs.
To develop a robust transmission policy for such an uncoordinated AP,
an agent, called the protagonist, learns through competition with a hypothetical adversarial agent during a training scenario.
It is worth noting that
we utilized a history of a set of actions of both agents, and designed a reward that depends on this set
to promote moderate competition and encourage the robustness of the policy.
The simulation results indicate that our proposed scheme leads to an improvement in the minimum sum of throughputs
of the protagonist and outsider.
However, with few transmission probabilities of the outsider, the sum of the throughputs does not improve;
therefore, learning a policy in a severe environment that was made by the adversary
leads to a more robust policy as the worst case under this scenario.

\section*{Acknowledgment}
This research and development work was supported by the MIC/SCOPE \#196000002.

\bibliographystyle{IEEEtran}
\bibliography{vtcf_2020}

\end{document}